\documentclass[aps,pra,showpacs,groupedaddress,twocolumn]{revtex4}
\usepackage{amsfonts}
\usepackage{graphicx}
\usepackage{latexsym}
\usepackage{makeidx}
\usepackage{amsmath}
\usepackage{amssymb}
\usepackage{graphicx}

\input{tcilatex}

\begin{document}

\title{Smooth quantum--classical transition in photon subtraction and
addition processes}
\author{A. V. Dodonov}
\affiliation{Departamento de F\'{\i}sica, Universidade Federal de S\~{a}o Carlos, P.O.
Box 676, S\~{a}o Carlos, 13565-905, S\~{a}o Paulo, Brazil}
\author{S. S. Mizrahi}
\affiliation{Departamento de F\'{\i}sica, Universidade Federal de S\~{a}o Carlos, P.O.
Box 676, S\~{a}o Carlos, 13565-905, S\~{a}o Paulo, Brazil}

\begin{abstract}
Recently Parigi \textit{et al.} [Science \textbf{317}, 1890 (2007)]
implemented experimentally the photon subtraction and addition processes
from/to a light field in a conditional way, when the required operations
were produced successfully only upon the positive outcome of a separate
measurement. It was verified that for a low intensity beam (quantum regime)
the bosonic annihilation operator $a$ does indeed describe a single photon
subtraction, while the creation operator $a^{\dagger }$ describes a photon
addition. Nonetheless, the exact formal expressions for these operations do
not always reduce to these simple identifications, and in this connection
here we deduce the general superoperators for multiple photons subtraction
and addition processes and analyze the statistics of the resulting states
for classical field states having an arbitrary intensity. We obtain closed
analytical expressions and verify that for classical fields with high
intensity (classical regime) the operators that describe photon subtraction
and addition processes deviate significantly from simply $a$ and $a^{\dagger
}$. Complementarily, we analyze in details such a smooth quantum-classical
transition as function of beam intensity for both processes.
\end{abstract}

\maketitle

\section{Introduction}

It is well known \cite{Gla130} that the probability for absorbing one photon
per unit time from an electromagnetic field is proportional to the average
value of the ordered product of the negative and positive frequency electric
field operators over some quantum state $\rho $, also known as the
statistical operator. In the simplest case of a single-mode, that
probability can be expressed in terms of the standard bosonic `annihilation'
and `creation' operators $a$ and $a^{\dagger }$, satisfying the commutation
relation $[a,a^{\dagger }]=1$, as%
\begin{equation}
P=\gamma \mathrm{Tr}\left( a\rho a^{\dagger }\right) ,  \label{p-ab}
\end{equation}%
where $\rho $ stands for the field state just before absorption\emph{\/} and
$\gamma $ is an appropriate coefficient. After interacting with some
detector that absorbs one photon, the field makes a transition\ to a new
state, which can be formally described by the action of a \emph{photon
subtraction superoperator} (PSS) $\mathcal{D}$ as \cite{SD}
\begin{equation}
\rho ^{\prime }=P^{-1}\mathcal{D}\rho ,  \label{def-J}
\end{equation}%
where $\rho ^{\prime }$ represents the field state immediately after the
subtraction of one photon and $P=\mathrm{Tr}(\mathcal{D}\rho )$ is the
probability for that process.

The hermiticity of $\rho ^{\prime }$ is always assured whenever $\mathcal{D}$
takes the form
\begin{equation}
\mathcal{D}\rho \equiv \gamma O\rho O^{\dagger }  \label{ge}
\end{equation}%
where $O$ is some\ operator responsible for the subtraction of one photon
from the field and the explicit form of $O$ depends on the details of the
field-detector interaction. Since the 1960s several models were proposed
\cite{KK64,Moll68,Imo90,Ag94} (see \cite{PL00} for more references therein)
and the first one \cite{SD} used the rather simple identification $O=a$,
when Eq. (\ref{ge}) becomes

\begin{equation}
\mathcal{A}\rho =\gamma a\rho a^{\dagger }  \label{J-A}
\end{equation}%
and we shall refer to it as \emph{A-model}. Although such a form, Eq. (\ref%
{J-A}), seems quite natural in view of equation (\ref{p-ab}), this choice
was, as a matter of fact, intuitive, although later, under certain
assumptions -- such as the weak coupling, low intensity and short
interaction time -- the authors of Refs. \cite{Imo90,DMD05} were able to
derive it from doing a microscopic\ analysis.

Nonetheless, if those assumptions are replaced by others, one can obtain
different superoperators $\mathcal{D}$. A family of PSS based on the \emph{%
nonlinear lowering operators}\ of the form $O=(1+\hat{n})^{-\beta }a$, where
$\hat{n}=a^{\dagger }a$, was derived in Ref. \cite{DMD05}. The special case $%
\beta =1/2$, called \emph{E-model},
\begin{equation}
\mathcal{E}\rho =E_{-}\rho E_{+},\quad E_{-}\equiv (1+\hat{n})^{-1/2}a
\label{J-E}
\end{equation}%
was originally proposed \emph{ad hoc} in \cite{benaryeh,OMD-JOB}. Later,
analyzing carefully a microscopic model, it was shown \cite{job05,DMD05,PRAS}
that both superoperators $\mathcal{A}$ and $\mathcal{E}$ are particular
cases of a more general one, where $\mathcal{A}$ ($\mathcal{E}$) is specific
for a small (large) mean photon number.

\begin{figure}[t]
\begin{center}
\includegraphics[width=.48\textwidth]{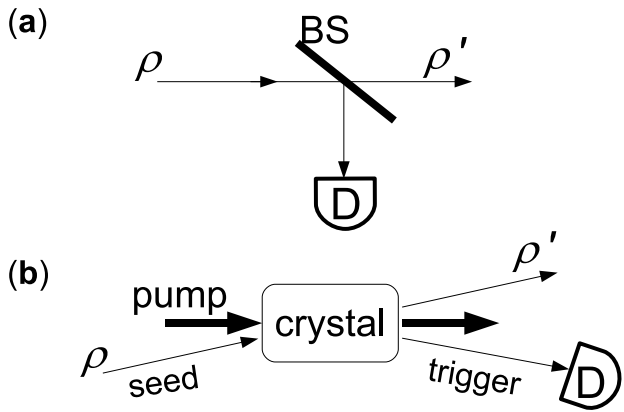} {}
\end{center}
\caption{(\textbf{a}) Scheme for conditional photon subtraction by means of
a beam-splitter (BS) and a photodetector (D). (\textbf{b}) Scheme for
conditional photon addition using a pump beam, a non-linear optical crystal
and a photodetector.}
\label{scheme}
\end{figure}

Recently, the A-model was subdued to an experimental verification in the low
photon number regime and weak field-detector coupling \cite{science}. That
experiment employed a beam-splitter and a single photon on/off detector
(SPD), within a simple scheme as illustrated in the Fig. \ref{scheme}a: a
low intensity beam, with a small mean photon number per unit time, is
prepared to hit a low reflectivity beam-splitter. The reflected beam is
continuously measured by the SPD, and whenever it clicks, the transmitted
field statistical operator can be described, approximately, by the A-model
\cite{science}.

However, as already observed in \cite{job05}, $\mathcal{A}$ is an unbounded
superoperator and some physical inconsistencies appear \cite{incon}. For
example, $\mathcal{A}$ does not hold for large values of the initial mean
photon number, since the probability of a photon subtraction, $P=\gamma
\mathrm{Tr}(a\rho a^{\dagger })$, may become larger than $1$. Moreover,
applying $\mathcal{A}$ on some specific field states, the calculations
predict results which look counterintuitive. For example, operating with $%
\mathcal{A}$ on the thermal state, which has the photon number distribution
\begin{equation}
p_{n}^{\left( th\right) }=\frac{n_{0}^{n}}{\left( n_{0}+1\right) ^{n+1}},
\label{thermal}
\end{equation}%
where $n_{0}$ is the mean photon number, for post-selected state we obtain $%
\left\langle n\right\rangle _{\mathcal{A}}=2n_{0}$. Although this prediction
was confirmed experimentally in the `quantum limit' of small mean photon
number, $n_{0}\lesssim 1$, it is counterintuitive from the classical point
of view, namely, a high intensity field cannot double the intensity due to a
one-photon detection. So $\mathcal{A}$ can describe satisfactorily the low
intensity field but not a high intensity one. Quite differently, using the
E-model we get $\langle n\rangle _{\mathcal{E}}=n_{0}$, which is sound for a
high intensity field after one-photon detection, whereas it does not
reproduce the observed doubling for low intensity. Therefore, it is
interesting to analyze the behavior of the PSS in the `classical limit'
(large mean photon number) and how the pattern of post-selected state
statistics changes smoothly and continuously as function of the field
intensity. The \emph{quantum-classical transition} in photon subtraction
process regards the passage between the two extreme situations for classical
field states that can be prepared with arbitrary low and high intensities.

Besides the photon subtraction, one may also implement the photon addition
operation by the conditional stimulated down-conversion in a non-linear
optical crystal, as reported in \cite{science} and illustrated in the Fig. %
\ref{scheme}b. Inside the crystal the pumped photons may decay spontaneously
into two entangled photons with energies that sum up to that of the parent
one. Upon detecting one of these photons (known as the trigger photon) along
a particular direction, the other photon state is unambiguously determined.
If one injects a seed light into the crystal, the stimulated emission may
occur into the same mode, and the detection of a single trigger photon (by
means of a SPD) indicates the conditional generation of the photon-added
state. Using this scheme, the one photon addition superoperator (PAS) for
low beam intensity is%
\begin{equation}
\mathbb{A}\rho =\gamma ^{\prime }a^{\dagger }\rho a
\end{equation}%
(we call it the \emph{A}$_{+}$\emph{-model}), which was experimentally
implemented for a low mean photon number \cite{science}. Since the
superoperator $\mathbb{A}$ is unbounded, with the probability of photon
addition becoming larger than $1$ for large field intensities, in analogy to
the E-model we also define the \emph{E}$_{+}$\emph{-model} as
\begin{equation}
\mathbb{E}\rho =E_{+}\rho E_{-},
\end{equation}%
that will also be analyzed and discussed below.

In the present paper we generalize the theoretical analysis presented in
\cite{science} for both, the photon subtraction and photon addition
operations, by substituting the SPD by the $k$-photon \emph{resolving} \cite%
{resol} or \emph{nonresolving} \cite{nonresol} detector, and obtain formally
PSS and PAS valid for an arbitrary field at any intensity. Although the
resulting formal expressions for the PSS and PAS are compact, in general, it
is not straightforward to obtain closed analytical expressions for these
operations for an arbitrary initial field state. Therefore, it is desirable
to have some simple approximate expressions for the PSS and PAS that hold in
a specific regime of parameters and can be easily evaluated for an arbitrary
field state, providing a simple means of predicting the outcome of the
experiment. In this connection, we show that in the quantum limit (small
photon number), the A-model and A$_{+}$-model are good approximations to the
exact PSS and PAS for any field state, respectively, while in the classical
limit (large photon number), the E- and E$_{+}$- models are more appropriate
for a nonresolving detector and the `\emph{classical}' field states:
coherent, thermal and `mixed light'.

The paper contains three additional sections. In section \ref{II} we obtain
exact analytical expressions for the PSS and analyze the behavior of the
photon subtraction probabilities and post-selected states statistics for any
intensity of the field and for different kinds of `classical' field states:
(a) coherent, (b) thermal and (c) `mixed light' \cite{Laguerre}, which, in
principle, can be produced in the laboratory with arbitrarily low or high
beam intensities. In section \ref{III} we do the same for the PAS,
considering (a) coherent and (b) thermal field states. In section \ref{IV}
we present a summary and our conclusions.

\section{Photon subtraction}

\label{II}

After passing through the beam-splitter a small fraction of the incident
field (signal) is reflected into the mode $b$, initially in the vacuum state
$|0_{b}\rangle \langle 0_{b}|$ (Fig. \ref{scheme}a). The post-selection
procedure -- in which $k>0$ photons are detected by a detector placed in the
reflected path -- allows to express the transmitted field state as Eq. (\ref%
{def-J}), where the exact PSS is written as
\begin{equation}
\mathcal{D}_{k}\rho =\mathrm{Tr}_{b}\left[ M_{k}U\left( \rho \otimes
|0_{b}\rangle \langle 0_{b}|\right) U^{\dagger }\right] ,
\end{equation}%
where $\rho $ is the incident field state. Here%
\begin{equation}
U=\exp \left[ \theta \left( a^{\dagger }b-ab^{\dagger }\right) \right]
\label{u}
\end{equation}%
is the beam-splitter operator, where $a$ (signal beam) and $b$ (reflected
beam) represent the two modes resulting from the incident beam, and $\theta $
is a parameter related to the reflectivity ($R=\sin ^{2}\theta $) and
transmittivity ($T=\cos ^{2}\theta $) coefficients. The operator $%
M_{k}=\sum_{l=k}^{\infty }\Upsilon _{l}|l_{b}\rangle \langle l_{b}|$, that
acts on the reflected mode, stands for the action of the detector (referred
as $\mathcal{D}k$, for short). $\Upsilon _{l}=1$ for a \emph{nonresolving} $k
$-photon detector ($N\mathcal{D}k$), that clicks whenever $k$ or more
photons are absorbed \cite{resol,nonresol}, and $\Upsilon _{l}=\delta _{l,k}$
for the \emph{resolving} $k$-photon detector ($R\mathcal{D}k$), that clicks
when exactly $k$ photons are absorbed \cite{resol}. The SPD corresponds to $N%
\mathcal{D}1$ according to this notation.

The resulting PSS is a generalization of the Eq. (\ref{ge})%
\begin{equation}
\mathcal{D}_{k}\rho =\sum_{l=k}^{\infty }\Upsilon _{l}\langle
l_{b}|U|0_{b}\rangle \rho \langle l_{b}|U|0_{b}\rangle ^{\dagger }.
\label{Dk}
\end{equation}%
The operator $U$ has the form $\exp \left[ -\theta \left( {K}_{+}+{K}%
_{-}\right) \right] $ where ${K}_{+}\equiv {b}^{\dagger }{a}$, ${K}%
_{-}\equiv -{b}{a}^{\dagger }$; a third operator, ${K}_{0}\equiv ({b}%
^{\dagger }{b}-{a}^{\dagger }{a})/2$, is necessary to close the su(1,1)
algebra, they constitute the generators of the SU(1,1) group. Using the
factorization theorem \cite{Ban93,Wei,alg,inter} we can write (\ref{Dk}) as%
\begin{equation}
\mathcal{D}_{k}\rho =\sum_{l=k}^{\infty }\Upsilon _{l}\frac{\left( \tan
^{2}\ \theta \right) ^{l}}{l!}a^{l}\left( \cos \theta \right) ^{\hat{n}}\rho
\left( \cos \theta \right) ^{\hat{n}}a^{\dagger l}.  \label{J}
\end{equation}%
which is the exact and complete PSS compatible with the experimental setup
reported in \cite{science}. For a small mean photon number $n_{0}$ and $R\ll
1$ one has roughly $\mathrm{Tr}\left[ \left( \cos \theta \right) ^{\hat{n}%
}\rho \left( \cos \theta \right) ^{\hat{n}}\right] \sim \left(
1-n_{0}R\right) $, so for $n_{0}R\ll 1$ one obtains%
\begin{equation}
\mathcal{D}_{k}\rho \simeq \mathcal{A}_{k}\rho \equiv \frac{R^{k}}{k!}%
a^{k}\rho a^{\dagger k}
\end{equation}%
for both kinds of detectors. For $k=1$ one retrieves the A-model, whereas
for $k>1$ we have a straightforward generalization of A-model for
multiphoton subtraction.

The photon number distribution of the post-selected state for the $k$
photons detector is
\begin{equation}
p_{n,\mathcal{D}k}^{\prime }=P_{\mathcal{D}k}^{-1}\Theta _{n}^{\left(
\mathcal{D}k\right) },
\end{equation}%
where%
\begin{equation}
\Theta _{n}^{\left( \mathcal{D}k\right) }=\sum_{l=k}^{\infty }\Upsilon _{l}%
\binom{n+l}{n}T^{n}R^{l}p_{n+l}
\end{equation}%
and $P_{\mathcal{D}k}=\sum_{n=0}^{\infty }\Theta _{n}^{\left( \mathcal{D}%
k\right) }$ is the probability for $k$-photon subtraction, with $p_{n}$
denoting the initial photon number distribution.\newline

Below we specialize to three different states of the field, which are
currently produced in the lab: the coherent, thermal and the `mixed light'
\cite{Laguerre}. The latter is a `mixture of the thermal and coherent
radiation' \cite{Lachs}, whose the photon number distribution is%
\begin{equation}
p_{n}^{\left( ml\right) }=\exp \left( -\frac{n_{c}}{1+n_{t}}\right) \frac{%
n_{t}^{n}}{\left( n_{t}+1\right) ^{n+1}}L_{n}\left[ -\frac{n_{c}}{%
n_{t}\left( 1+n_{t}\right) }\right] .
\end{equation}%
where $n_{c}$ and $n_{t}$ are the mean photon numbers of the coherent and
thermal part of the mixed light, respectively, $n_{0}=n_{c}+n_{t}$ is the
total photon number and $L_{n}(\cdot )$ is a Laguerre polynomial. For these
three states the resulting expressions for $p_{n,\mathcal{D}k}^{\prime }$, $%
P_{\mathcal{D}k}$ and the two lower moments of the photon number
distribution for the post-selected state are given in the appendix \ref{A}.

\begin{figure}[t]
\begin{center}
\includegraphics[width=.5\textwidth]{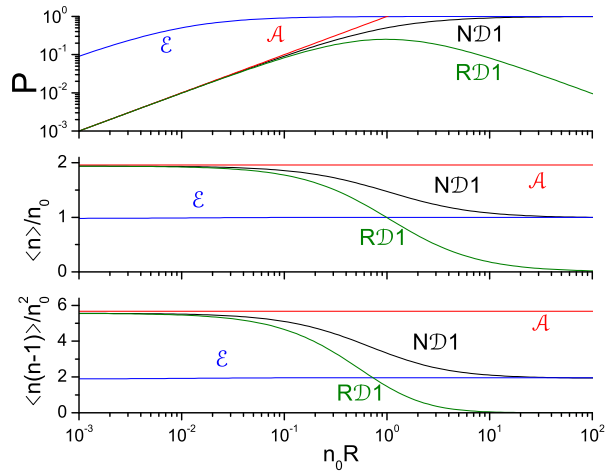} {}
\end{center}
\caption{(Color online) $N\mathcal{D}1$ and $R\mathcal{D}1$ results for the
`mixed light' state with mean photon number $n_{0}$ compared to the
predictions given by superoperators $\mathcal{A}$ and $\mathcal{E}$. Here $%
n_{c}=n_{t}/4$.}
\label{1}
\end{figure}

\begin{figure}[t]
\begin{center}
\includegraphics[width=.5\textwidth]{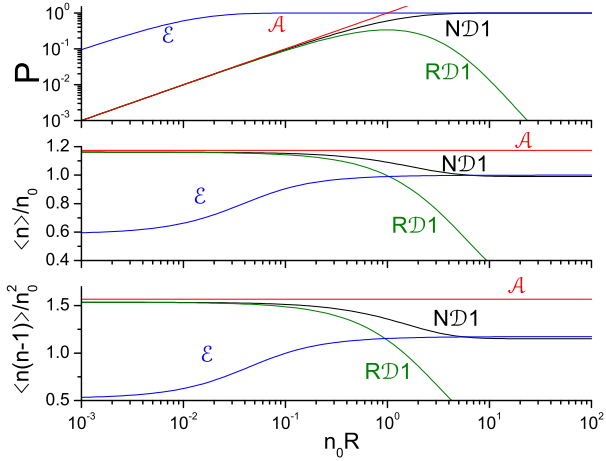} {}
\end{center}
\caption{(Color online) Same as Fig. \protect\ref{1} for $n_{c}=10n_{t}$.}
\label{2}
\end{figure}

For a small mean photon number, $n_{0}R\ll 1$, the expressions for the
photon subtraction probability and the two lower moments of photon number in
the resulting state [see Eqs. in appendix \ref{A}] are approximately equal
to the expressions corresponding to the generalized A-model. Thus, in the
quantum regime (low intensity field), $\mathcal{A}_{k}$ is a good
approximation for the PSS, and the detection of one or more photons may
significantly increase the mean photon number of the post-selected state. On
the other hand, for a high intensity field, $n_{0}R\gg 1$, the results are
quite different from those predicted by using $\mathcal{A}_{k}$. This
behavior can be appreciated looking Figs. \ref{1} and \ref{2}, where we have
plot $P_{\mathcal{D}1}$, $\left\langle n\right\rangle _{\mathcal{D}1}$ and $%
\left\langle n\left( n-1\right) \right\rangle _{\mathcal{D}1}$, together
with the corresponding predictions of $\mathcal{A}_{1}$ and $\mathcal{E}_{1}$%
, for the mixed light state with $n_{c}=n_{t}/4$ (Fig. \ref{1}) and $%
n_{c}=10n_{t}$ (Fig. \ref{2}), setting $R=10^{-2}$. One can perceive, from
the figures, that for $n_{0}R\gg 1$ the probabilities and the factorial
moments for $N\mathcal{D}1$ are close to those predicted by the generalized
E-model. Therefore, for the considered classical field states, in the
classical regime $\mathcal{E}_{k}$ represents better the photon subtraction
process (using a nonresolving detector) and, as verified empirically, the
mean photon number does not increase upon the photodetection. The smooth
transition between the A- and E- models occurs in the region $n_{0}R\sim 1$.

The important effect of the measurement back-action may be clearly verified
by comparing the outcomes of the instantaneous $k$-photon detection (using
either $N\mathcal{D}k$ or $R\mathcal{D}k$) to the sequential detection of $k$
photons (one by one, as a sequence of discernible clicks), using an array of
$k$ SPD's and detecting one click in each one. For the sequential $k$
photons detection the PSS is $\mathcal{S}_{k}=\left( \mathcal{D}_{N1}\right)
^{k},$where $\mathcal{D}_{N1}$ is given by Eq. (\ref{J}) with $\Upsilon
_{l}=1$. The resulting expressions are given in the appendix \ref{C} and
plotted in the Fig. \ref{3}. Looking at it we can compare the outcomes of
the three kinds of detectors, $\mathcal{S}2$, $N\mathcal{D}2$ and $R\mathcal{%
D}2$, for the thermal state and $R=10^{-2}$. It turns out that for a
sequential counting, the probability of detecting $k$ photons is always
higher than that calculated by admitting an instantaneous detection.
Moreover, for $n_{0}R\ll 1$ the mean photon number in the post-selected
state is always higher than for instantaneous detection, while for $%
n_{0}R\gtrsim 1$ different mean values are predicted, such that $%
\left\langle n\right\rangle _{N\mathcal{D}k}>\left\langle n\right\rangle _{%
\mathcal{S}k}>\left\langle n\right\rangle _{R\mathcal{D}k}$, see Fig. \ref{3}%
.
\begin{figure}[t]
\begin{center}
\includegraphics[width=.5\textwidth]{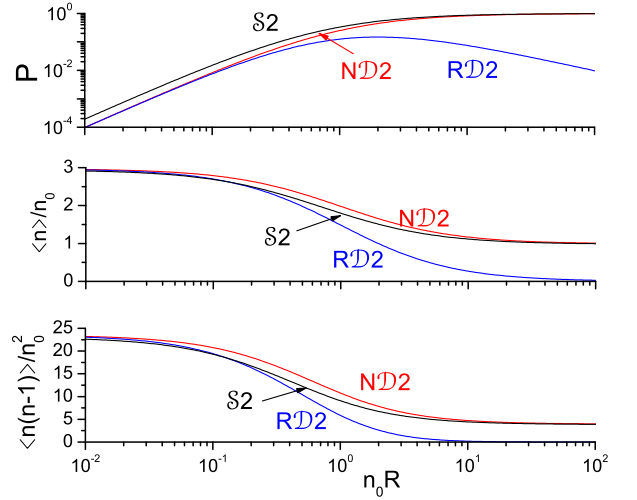} {}
\end{center}
\caption{(Color online) Comparison of 2 sequential detections with SPD's to
the 2-photon detection using either the resolving or the nonresolving
detectors, for the thermal state.}
\label{3}
\end{figure}

\section{Photon addition}

\label{III}

The $k$-photon addition superoperator is defined as \cite{science}%
\begin{equation}
\mathbb{D}_{k}\rho =\mathrm{Tr}_{b}\left[ M_{k}u\rho |0_{b}\rangle \langle
0_{b}|u^{\dagger }\right] ,  \label{drho1}
\end{equation}%
where $M_{k}=\sum_{l=k}^{\infty }\Upsilon _{l}|l_{b}\rangle \langle l_{b}|$
(as defined previously) and
\begin{equation}
u=\exp \left[ \lambda \left( ab-a^{\dagger }b^{\dagger }\right) \right]
\end{equation}%
is the operator describing the parametric down-conversion process with gain
factor $\lambda \ll 1$. The operator $u$ contains two generators of the $%
su(1,1)$ algebra, $\tilde{K}_{+}=a^{\dagger }b^{\dagger }$, $\tilde{K}_{-}=ab
$, while the third one is $\tilde{K}_{0}=\left( a^{\dagger }a+b^{\dagger
}b+1\right) /2$. So Eq. (\ref{drho1}) becomes \cite{inter}%
\begin{equation}
\mathbb{D}_{k}\rho =t\sum_{l=k}^{\infty }\Upsilon _{l}\frac{\left( \tanh
^{2}\lambda \right) ^{l}}{l!}a^{\dagger l}\left( \cosh ^{-1}\lambda \right)
^{\hat{n}}\rho \left( \cosh ^{-1}\lambda \right) ^{\hat{n}}a^{l},  \label{J1}
\end{equation}%
and we define $r\equiv \sinh ^{2}\lambda $ and $t\equiv \cosh ^{-2}\lambda $%
. For a small mean photon number one obtains%
\begin{equation}
\mathbb{D}_{k}\rho \simeq \mathbb{A}_{k}\rho =\frac{r^{k}}{k!}a^{\dagger
k}\rho a^{k},
\end{equation}%
which is the generalization of the A$_{+}$-model. As like as $\mathcal{A}_{k}
$, the superoperator $\mathbb{A}_{k}$ is unbounded, with the probability of
photon addition becoming larger than $1$ for $n_{0}\gg 1$. Therefore, the A$%
_{+}$-model cannot stand for high field intensities. Here we shall study the
behavior of the lowest photon number moments in the transition from the
quantum regime, $n_{0}\ll 1$, to the classical one, $n_{0}\gg 1$, and
compare the results with the predictions of the generalized E$_{+}$-model, $%
\mathbb{E}_{k}\rho =E_{+}^{k}\rho E_{-}^{k}$. Contrarily to $\mathbb{A}_{k}$%
, superoperator $\mathbb{E}_{k}$ is bounded, presenting probability $P_{%
\mathbb{E}k}=1$ and the mean value $\left\langle n\right\rangle _{\mathbb{E}%
k}=n_{0}+k$ for $k$-photon addition operation.

The photon number distribution of the $k$-photon added state is given by
\begin{equation}
p_{n,\mathbb{D}k}^{\prime }=P_{\mathbb{D}k}^{-1}\Theta _{n}^{\left( \mathbb{D%
}k\right) },
\end{equation}%
where%
\begin{equation}
\Theta _{n}^{\left( \mathbb{D}k\right) }=\sum_{l=k}^{\infty }\Upsilon _{l}%
\binom{n}{l}t^{n+1}r^{l}p_{n-l}
\end{equation}%
and the probability of the $k$-photon addition is $P_{\mathbb{D}%
k}=\sum_{n=0}^{\infty }\Theta _{n}^{\left( \mathbb{D}k\right) }$,
independently of the number of photons present in the field. In order to
make the physics more transparent, we assume the simplest situation $k=1$,
for which $P_{\mathbb{A}1}=r\left( n_{0}+1\right) $ and $\left\langle
n\right\rangle _{\mathbb{E}1}/n_{0}=1+n_{0}^{-1}$.

\begin{figure}[t]
\begin{center}
\includegraphics[width=.5\textwidth]{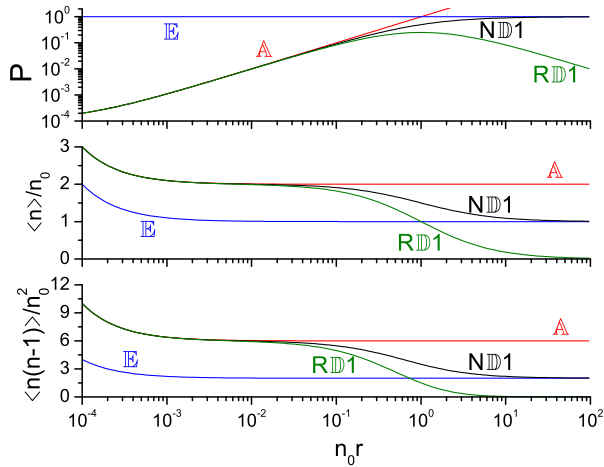} {}
\end{center}
\caption{(Color online) $N\mathbb{D}1$ and $R\mathbb{D}1$ results for the
thermal state with mean photon number $n_{0}$ compared to the predictions of
superoperators $\mathbb{A}$ and $\mathbb{E}$.}
\label{4}
\end{figure}

For the two different field states, the coherent and the thermal, we get the
expressions given in the appendix \ref{B}. For $n_{0}r\ll 1$, in both cases,
the expressions become very close to those predicted by the A$_{+}$-model,
so in the quantum regime the photon creation operator $a^{\dagger }$
describes accurately the photon addition process. In the regime $n_{0}r\gg 1$%
, for the coherent state the predictions of the A$_{+}$- and E$_{+}$- models
for the moments $\left\langle n\right\rangle $ and $\left\langle n\left(
n-1\right) \right\rangle $ are very similar, while for the thermal state,
according to the A$_{+}$-model, the mean photon number is roughly twice the
one predicted by the E$_{+}$-model. In Fig. \ref{4} we show the behaviors of
the photon addition probability $P$, $\left\langle n\right\rangle $ and $%
\left\langle n\left( n-1\right) \right\rangle $ for the thermal state as
function of $n_{0}r$ for $\lambda =10^{-2}$. We see that by using a
nonresolving detector for $n_{0}r\gg 1$, the expressions approach the
results of the E$_{+}$-model, so in the classical regime the A$_{+}$-model
ceases to represent the PAS and the E$_{+}$-model becomes more appropriate.
The transition between the A$_{+}$- and E$_{+}$- models occurs in the region
$n_{0}r\sim 1$. Finally, notice that, although the photon resolving and
nonresolving detectors are quite similar in the quantum regime, they are
completely different in the classical one, as clearly seen from Figs. \ref{1}%
, \ref{2}, \ref{3} and \ref{4}.

\section{Summary and conclusions}

\label{IV}

In summary, we have analyzed the quantum--classical transitions for the
photon subtraction and for the photon addition processes from or into the
field. We used the usual formal representations for the beam-splitter, the
optical nonlinear crystal and the $k$-photon resolving or nonresolving
detector, as it was described in \cite{science}. We considered three \emph{%
classical} field states -- coherent, thermal and the `mixed light' -- to
illustrate the transition from quantum to classical regimes and studied the
photon number factorial moments in the post-selected state as function of
the intensity of the prepared field, that goes on either a beam-splitter or
a nonlinear crystal. We obtained the formal expressions for the PSS and the
PAS, valid for any input state, and for the considered classical states we
derived closed analytical expressions for the photon number distribution of
the post-selected state and the associated lower factorial moments, as well
as the photon addition and subtraction probabilities.

We found that in the quantum regime (small photon number) the PSS can be
described approximately by the generalized A-model, while the PAS by the
generalized\thinspace A$_{+}$-model for any field state. The mean photon
number in the post-selected state may increase significantly due to the
photon subtraction. On the other hand, in the classical regime (large photon
number) the generalized A- and A$_{+}$- models lose the validity and, for a
nonresolving photodetector and the considered classical states, the PSS
(PAS) is better approximated by the generalized E-model (E$_{+}$-model), and
the mean photon number necessarily decreases upon a photon subtraction.
Thus, one may associate $E_{-}$ used in \cite{OMD-JOB} to the classical
photodetection operator, in the same way as the bosonic annihilation
operator $a$ is associated to the quantum photodetection operator.

In conclusion, the exact expressions of photon subtraction and addition
superoperators, Eqs. (\ref{J}) and (\ref{J1}), cannot always be written as $%
\gamma O\rho O^{\dagger }$ or $\gamma O^{\dagger }\rho O$, they reduce to
these simple expressions only in some limits of the field intensity or for a
photon-number resolving detector. More importantly, they depend on the form
how detection is done, so the post-selected field state will depend
essentially on the way the experimenter chooses to probe it.

\begin{acknowledgments}
SSM acknowledges financial support from CNPq and FAPESP, brazilian agencies.
AVD would like to thank, in particular, the FAPESP, Grant No. 04/13705-3.
\end{acknowledgments}

\appendix

\section{Expressions for photon subtraction}

\label{A}

\textbf{(a)} The coherent state $|\alpha \rangle $ is insensitive to the
detector outcome, since it is an eigenstate of the lowering operator $a$
appearing in Eq. (\ref{u}), so the resulting post-selected state is still a
coherent state $|\alpha \cos \theta \rangle $, although having the field
intensity attenuated. We get for the photon number distribution of the
post-selected state%
\begin{equation}
p_{n,N\mathcal{D}k}^{\prime }=p_{n,R\mathcal{D}k}^{\prime
}=e^{-n_{0}T}\left( n_{0}T\right) ^{n}/n!.
\end{equation}%
The normalized factorial moments in the post-selected state are the same for
both detectors,
\begin{equation}
\left\langle n\right\rangle _{\mathcal{D}k}/n_{0}=T,\quad \left\langle
n\left( n-1\right) \right\rangle _{\mathcal{D}k}/n_{0}^{2}=T^{2},
\end{equation}%
although the subtraction probabilities associated to the detectors are
different%
\begin{equation}
P_{R\mathcal{D}k}=e^{-n_{0}R}\frac{\left( n_{0}R\right) ^{k}}{k!},\quad P_{N%
\mathcal{D}k}=1-e^{-n_{0}R}\sum_{l=0}^{k-1}\frac{\left( n_{0}R\right) ^{l}}{%
l!}.
\end{equation}

\textbf{(b)} For the thermal state the photon number distributions of the
post-selected state for the $k$ photons resolving/nonresolving detector are%
\begin{equation}
p_{n,R\mathcal{D}k}^{\prime }=\binom{n+k}{n}\left( \frac{1+n_{0}R}{1+n_{0}}%
\right) ^{k+1}\left( \frac{n_{0}T}{1+n_{0}}\right) ^{n}
\end{equation}%
\begin{eqnarray}
p_{n,N\mathcal{D}k}^{\prime } &=&\left( \frac{1+n_{0}R}{n_{0}R}\right) ^{k}%
\left[ \frac{\left( n_{0}T\right) ^{n}}{\left( 1+n_{0}T\right) ^{n+1}}%
\right.  \\
&&\left. -\sum_{l=0}^{k-1}\binom{n+l}{n}\left( \frac{n_{0}R}{1+n_{0}}\right)
^{l}\frac{\left( n_{0}T\right) ^{n}}{\left( 1+n_{0}\right) ^{n+1}}\right] .
\notag
\end{eqnarray}%
The $k$-photon subtraction probability is%
\begin{equation}
P_{\mathcal{D}k}=\frac{\left( n_{0}R\right) ^{k}}{\left( 1+n_{0}R\right)
^{k+1-v}},
\end{equation}%
and the normalized lowest factorial moments are%
\begin{equation}
\frac{\left\langle n\right\rangle _{\mathcal{D}k}}{n_{0}}=T\frac{1+k+vn_{0}R%
}{1+n_{0}R}
\end{equation}%
\begin{equation}
\frac{\left\langle n\left( n-1\right) \right\rangle _{\mathcal{D}k}}{%
n_{0}^{2}}=T^{2}\frac{\left( 1+k\right) \left( 2+k\right) +2vn_{0}R\left(
2+k+n_{0}R\right) }{\left( 1+n_{0}R\right) ^{2}},
\end{equation}%
where $v=1$ for the $N\mathcal{D}k$ and $v=0$ for the $R\mathcal{D}k$.

\textbf{(c)} For the `mixed light' and the nonresolving detector $N\mathcal{D%
}k$, the resulting expressions are quite extensive, so we write out
explicitly only the expressions for $k=1$,%
\begin{equation}
P_{N\mathcal{D}1}=1-\frac{e^{-x}}{1+Rn_{t}}
\end{equation}%
\begin{equation}
\left\langle n\right\rangle _{N\mathcal{D}1}=P_{N\mathcal{D}1}^{-1}T\left[
n_{0}-e^{-x}\frac{n_{0}+Rn_{t}^{2}}{\left( 1+Rn_{t}\right) ^{3}}\right]
\end{equation}%
\begin{gather}
\left\langle n\left( n-1\right) \right\rangle _{N\mathcal{D}1}=P_{N\mathcal{D%
}1}^{-1}T^{2}\left\{ n_{0}^{2}+n_{t}\left( n_{0}+n_{c}\right) \right.  \\
-\left. e^{-x}\frac{n_{0}^{2}+2n_{c}n_{t}+n_{t}^{2}\left(
1+4Rn_{0}+2R^{2}n_{t}^{2}\right) }{\left( 1+Rn_{t}\right) ^{5}}\right\} ,
\notag
\end{gather}%
where $x\equiv Rn_{c}/\left( 1+Rn_{t}\right) $. For the $R\mathcal{D}k$ the
corresponding expressions are quite simple for any $k$%
\begin{equation}
P_{R\mathcal{D}k}=e^{-x}\frac{\left( Rn_{t}\right) ^{k}}{\left(
1+Rn_{t}\right) ^{k+1}}L_{k}
\end{equation}%
\begin{equation}
\left\langle n\right\rangle _{R\mathcal{D}k}=\frac{n_{t}T}{1+Rn_{t}}\left[
z-k\frac{L_{k-1}}{L_{k}}\right]
\end{equation}%
\begin{eqnarray}
\left\langle n\left( n-1\right) \right\rangle _{R\mathcal{D}k} &=&\left(
\frac{n_{t}T}{1+Rn_{t}}\right) ^{2}\left\{ z\left( 1+z\right) -y\right.  \\
&-&\left. \frac{2kzL_{k-1}}{L_{k}}+\frac{k\left( k-1\right) L_{k-2}}{L_{k}}%
\right\}   \notag
\end{eqnarray}%
where $L_{k}\equiv L_{k}\left( y\right) $, $y\equiv -n_{c}/\left[
n_{t}\left( 1+Rn_{t}\right) \right] $ and $z\equiv 1+2k-y$.

\section{Expressions for sequential photon subtraction}

\label{C}

For the sequential counting, in the simplest case, $k=2$, we obtain:\newline
\textbf{(a)} For a coherent state $p_{n,\mathcal{S}2}^{\prime
}=e^{-n_{0}T^{2}}\left( n_{0}T^{2}\right) ^{n}/n!$ and
\begin{gather}
P_{\mathcal{S}2}=e^{-n_{0}R\left( 1+T\right) }\left( e^{n_{0}RT}-1\right)
\left( e^{n_{0}R}-1\right)   \notag \\
\frac{\left\langle n\right\rangle _{\mathcal{S}2}}{n_{0}}=T^{2},\quad \frac{%
\left\langle n\left( n-1\right) \right\rangle _{\mathcal{S}2}}{n_{0}^{2}}%
=T^{4}.
\end{gather}%
\textbf{(b)} For a thermal state%
\begin{eqnarray}
p_{n,\mathcal{S}2}^{\prime } &=&g_{1}^{-1}\left[ \frac{\left(
n_{0}T^{2}\right) ^{n}}{\left( 1+n_{0}T^{2}\right) ^{n+1}}-\frac{\left(
n_{0}T^{2}\right) ^{n}}{\left( 1+n_{0}T\right) ^{n+1}}\right.  \\
&&\left. -\frac{\left( n_{0}T^{2}\right) ^{n}}{\left[ 1+n_{0}\left(
1-RT\right) \right] ^{n+1}}+\frac{\left( n_{0}T^{2}\right) ^{n}}{\left(
1+n_{0}\right) ^{n+1}}\right]   \notag
\end{eqnarray}%
and%
\begin{equation}
P_{\mathcal{S}2}=g_{1},\quad \frac{\left\langle n\right\rangle _{\mathcal{S}%
2}}{n_{0}}=T^{2}\,\frac{g_{2}}{g_{1}},\quad \frac{\left\langle n\left(
n-1\right) \right\rangle _{\mathcal{S}2}}{n_{0}^{2}}=2T^{4}\,\frac{g_{3}}{%
g_{1}},
\end{equation}%
where%
\begin{equation}
g_{n}\equiv 1-\left[ 1+n_{0}R\right] ^{-n}-\left[ 1+n_{0}RT\right] ^{-n}+%
\left[ 1+n_{0}R\left( 1+T\right) \right] ^{-n}.
\end{equation}

These expressions are quite different from the corresponding expressions
obtained in the appendix \ref{A}.

\section{Expressions for photon addition}

\label{B}

(\textbf{a)} For a coherent state and the one-photon nonresolving detector
we get%
\begin{equation}
p_{n,N\mathbb{D}1}^{\prime }=P_{N\mathbb{D}1}^{-1}\,te^{-n_{0}}\left[ \left(
rt\right) ^{n}L_{n}\left( -\frac{n_{0}}{r}\right) -\frac{\left(
n_{0}t\right) ^{n}}{n!}\right]
\end{equation}%
\begin{equation}
P_{N\mathbb{D}1}=1-te^{-rtn_{0}}
\end{equation}%
\begin{equation}
\frac{\left\langle n\right\rangle _{N\mathbb{D}1}}{n_{0}}=\frac{%
1+r+r/n_{0}-t^{2}e^{-rtn_{0}}}{P_{N\mathbb{D}1}}
\end{equation}%
\begin{equation}
\frac{\left\langle n\left( n-1\right) \right\rangle _{N\mathbb{D}1}}{%
n_{0}^{2}}=\frac{t^{-2}+4r/(n_{0}t)+2r^{2}/n_{0}^{2}-t^{3}e^{-rtn_{0}}}{P_{N%
\mathbb{D}1}},
\end{equation}%
where $L_{n}(\cdot )$ is a Laguerre polynomial. For the A$_{+}$- and E$_{+}$%
- models the resulting expressions are%
%
\begin{eqnarray}
&\left\langle n\right\rangle _{\mathbb{A}1}/n_{0}=1+n_{0}^{-1}+\left(
n_{0}+1\right) ^{-1}& \\
&\left\langle n\left( n-1\right) \right\rangle _{\mathbb{A}%
1}/n_{0}^{2}=1+4n_{0}^{-1}& \\
&\left\langle n\left( n-1\right) \right\rangle _{\mathbb{E}%
1}/n_{0}^{2}=1+2n_{0}^{-1}&.
\end{eqnarray}

(\textbf{b)} For a thermal state and the one-photon nonresolving detector
one has%
\begin{equation}
p_{n,N\mathbb{D}1}^{\prime }=P_{N\mathbb{D}1}^{-1}\,t\left[ \frac{\left(
rt+n_{0}\right) ^{n}}{\left( 1+n_{0}\right) ^{n+1}}-\frac{\left(
n_{0}t\right) ^{n}}{\left( 1+n_{0}\right) ^{n+1}}\right]
\end{equation}%
\begin{equation}
P_{N\mathbb{D}1}=\frac{rt\left( 1+n_{0}\right) }{1+n_{0}rt}
\end{equation}%
\begin{equation}
\frac{\left\langle n\right\rangle _{N\mathbb{D}1}}{n_{0}}=\frac{\left(
t^{-1}+r/n_{0}\right) -\left[ t/\left( n_{0}rt+1\right) \right] ^{2}}{P_{N%
\mathbb{D}1}}
\end{equation}%
\begin{equation}
\frac{\left\langle n(n-1)\right\rangle _{N\mathbb{D}1}}{n_{0}^{2}}=2\frac{%
\left( t^{-1}+r/n_{0}\right) ^{2}-\left[ t/\left( n_{0}rt+1\right) \right]
^{3}}{P_{N\mathbb{D}1}}.
\end{equation}%
For the one-photon resolving detector we get%
%
\begin{eqnarray}
p_{n,R\mathbb{D}1}^{\prime }&=&\frac{\left( 1+n_{0}rt\right) ^{2}}{t\,n_{0}}\,n%
\frac{\left( n_{0}t\right) ^{n}}{\left( 1+n_{0}\right) ^{n+1}} \\
P_{R\mathbb{D}1}&=&\frac{rt^{2}\left( 1+n_{0}\right) }{\left( 1+n_{0}rt\right)
^{2}}\\
\frac{\left\langle n\right\rangle _{R\mathbb{D}1}}{n_{0}}&=&\frac{%
1+t+n_{0}^{-1}}{1+n_{0}rt}\\
\frac{\left\langle n(n-1)\right\rangle _{R\mathbb{D}1}}{n_{0}^{2}}&=&2t\frac{%
2+t+2n_{0}^{-1}}{\left( 1+n_{0}rt\right) ^{2}}\,.
\end{eqnarray}%
The expressions for the A$_{+}$- and E$_{+}$- models are%
%
\begin{eqnarray}
\left\langle n\right\rangle _{\mathbb{A}1}/n_{0}&=&2+n_{0}^{-1} \\
\left\langle n\left( n-1\right) \right\rangle _{\mathbb{A}%
1}/n_{0}^{2}&=&6+4n_{0}^{-1} \\
\left\langle n\left( n-1\right) \right\rangle _{\mathbb{E}%
1}/n_{0}^{2}&=&2+2n_{0}^{-1}.
\end{eqnarray}


\begin{thebibliography}{99}
\bibitem{incon} For $\mathcal{A}$ they were noticed already in the original
paper \cite{SD}; see also \cite{OMD-JOB,job05}.
\end{thebibliography}
\end{document}